\documentclass[epj]{webofc}
\usepackage[utf8]{inputenc}
\usepackage[varg]{txfonts}   
\usepackage{booktabs}
\usepackage{xcolor}
\definecolor{darkred}{rgb}{0.4,0.0,0.0}
\definecolor{darkgreen}{rgb}{0.0,0.4,0.0}
\definecolor{darkblue}{rgb}{0.0,0.0,0.4}
\usepackage[bookmarks,linktocpage,colorlinks,
linkcolor = darkred,
urlcolor  = darkblue,
citecolor = darkgreen]{hyperref}

\usepackage{subfigure}

\usepackage{caption}
\wocname{EPJ Web of Conferences}
\woctitle{Lattice2017}

\begin{document}

\selectlanguage{english}

\title{
  Application of tensor network method \\ to two dimensional lattice $\mathcal{N}=1$ Wess--Zumino model
}

\author{
  \firstname{Ryo}  \lastname{Sakai}\inst{1}\fnsep\thanks{Speaker, \email{sakai@hep.s.kanazawa-u.ac.jp}} \and
  \firstname{Daisuke} \lastname{Kadoh}\inst{2} \and
  \firstname{Yoshinobu} \lastname{Kuramashi}\inst{3,4} \and
  \firstname{Yoshifumi}  \lastname{Nakamura}\inst{4} \and
  \firstname{Shinji}  \lastname{Takeda}\inst{1} \and
  \firstname{Yusuke}  \lastname{Yoshimura}\inst{3}
}

\institute{
  Institute for Theoretical Physics, Kanazawa University, Kanazawa 920-1192, Japan
  \and
  Research and Educational Center for Natural Sciences, Keio University, Yokohama 223-8521, Japan
  \and
  Center for Computational Sciences, University of Tsukuba, Tsukuba 305-8577, Japan
  \and
  RIKEN Advanced Institute for Computational Science, Kobe 650-0047, Japan
}

\abstract{
  We study a tensor network formulation of the two dimensional lattice $\mathcal{N}=1$ Wess--Zumino model with Wilson derivatives for both fermions and bosons.
  The tensor renormalization group allows us to compute the partition function without the sign problem, and basic ideas to obtain a tensor network for both fermion and scalar boson systems were already given in previous works.
  In addition to improving the methods, we have constructed a tensor network representation of the model including the Yukawa-type interaction of Majorana fermions and real scalar bosons.
  We present some numerical results.
}

\maketitle

\section{Introduction}
\label{sec:Introduction}

In addition to the phenomenological expectations, supersymmetry attracts also the theoretical interests, for example superstring theory, AdS/CFT correspondence, and Seiberg--Witten theory.
As in many other cases, lattice studies are often needed in supersymmetric theories to analyze non-perturbative effects or to confirm theoretical conjectures.
However, despite the strong motivation of non-perturbative treatment, one cannot simply apply Monte Carlo simulations to supersymmetric lattice field theories owing to the sign problem.
The $\mathcal{N}=1$ Wess--Zumino model~\cite{Wess:1974tw,Ferrara:1975nf} in two dimensions is one of the simplest supersymmetric model and suffering from the sign problem on the lattice.
Even though some numerical approaches have been already attempted and reported in Refs.~\cite{Catterall:2003ae,Wozar:2011gu,Steinhauer:2014yaa} for this model, development of non-stochastic methods remains to be an important issue.

In this study, we apply the tensor renormalization group (TRG)~\cite{Levin:2006jai} to the model.
The TRG is a deterministic coarse-graining algorithm for the tensor network and completely free of the sign problem.
Once a partition function or Green's functions are represented as a tensor network, it can be computed in the TRG scheme.
Since Levin and Nave introduced the TRG in a two dimensional classical spin system, its validity has been shown for some two dimensional quantum field theories, \textit{e.g.} the $\phi^{4}$ model~\cite{Shimizu:2012wfa}, the Schwinger model~\cite{Shimizu:2014uva,Shimizu:2014fsa}, the $N_{\mathrm{f}}=1$ Gross--Neveu model~\cite{Takeda:2014vwa}, and the $\mathrm{CP}\left(N-1\right)$ model~\cite{Kawauchi:2016xng}.
Then, one is ready to tackle the supersymmetric models which consists of bosons and fermions using the TRG.

In this report we present a tensor network representation for the model which has Wilson derivatives for both fermions and bosons.
In Sec.~\ref{sec:Model} we introduce the model and discuss the tensor network representation with a focus on the fermion and the boson part in turn, and numerical results for the free case are given in Sec.~\ref{sec:Results}.

\section{Tensor network representation}
\label{sec:Model}

\subsection{Two dimensional lattice $\mathcal{N}=1$ Wess--Zumino model}

The Euclidean Lagrangian density of the two dimensional $\mathcal{N}=1$ Wess--Zumino model is defined by
\begin{align}
  \label{eq:2}
  \mathcal{L}^{\mathrm{Cont.}}
  = \frac{1}{2}\left(\partial_{\mu}\phi\right)^{2}
  + \frac{1}{2}\bar{\psi}\left(\partial\hspace{-1.5mm}{/}
  + P^{\prime}\left(\phi\right)\right)\psi
  + \frac{1}{2}P^{}\left(\phi\right)^{2},
\end{align}
where $\phi$, $\psi$, and $P\left(\phi\right)$ denote real scalar field, a two component Majorana spinor field, and the derivative of the superpotential $W\left(\phi\right)$ (\textit{i.e.} $P\left(\phi\right)=W^{\prime}\left(\phi\right)$), respectively.
Majorana spinors satisfy 
\begin{align}
  \label{eq:23}
  \psi=\psi^{\mathrm{c}}=C\bar{\psi}^{\mathrm{T}},
\end{align}
where $C$ is the charge conjugation matrix,
and $C$ and $\gamma_{\mu}$ obey the relations
\begin{align}
  \label{eq:24}
  C^{\mathrm{T}}=-C,\quad C^{\dagger}=C^{-1},\quad C^{-1}\gamma_{\mu}C=-\gamma_{\mu}^{\mathrm{T}},\quad\gamma_{\mu}\gamma_{\nu}+\gamma_{\nu}\gamma_{\mu}=2\delta_{\mu\nu}.
\end{align}
When we adopt the Wilson-type discretization for the fermion part of this model, we need to include the Wilson derivative also for the boson part to guarantee the supersymmetry restoration in the continuum limit~\cite{Bartels:1982ue,Golterman:1988ta}.
Therefore, the lattice action is given by
\begin{align}
  \label{eq:3}
  S
  = \sum_{n}\left[\frac{1}{2}\left(\partial_{\mu}^{\mathrm{s}}\phi\right)_{n}^{2} + \frac{1}{2}\bar{\psi}_{n}\left(D\psi\right)_{n} + \frac{1}{2}\left\{-\frac{r}{2}\left(\partial_{\mu}^{*}\partial_{\mu}^{}\phi\right)_{n} + P^{}\left(\phi_{n}\right)\right\}^{2}\right],
\end{align}
where $n=\left(n_{1}, n_{2}\right)$ represents the lattice coordinate in two dimensions,
and $D$ is the Wilson--Dirac operator
\begin{align}
  \label{eq:4}
  D
  = \partial\hspace{-1.5mm}{/}^{\mathrm{s}}
  - \frac{r}{2}\partial_{\mu}^{*}\partial_{\mu}^{}
  + P^{\prime}\left(\phi\right)
\end{align}
with the Wilson parameter $r$.
In this report, lattice units $a = 1$ are assumed,
and the forward, backward, and symmetric lattice derivatives are represented as $\partial_{\mu}$, $\partial^{*}_{\mu}$, and $\partial^{\mathrm{s}}_{\mu}$, respectively.
In the absence of interactions, the lattice action~(\ref{eq:3}) is invariant under the supersymmetric transformation:
\begin{align}
  \label{eq:5}
  &\delta \phi_{n}
    = \bar{\epsilon}\psi_{n}, \\
  &\delta \psi_{n}
    = \left[\left(\gamma_{\mu}\partial^{\mathrm{s}}_{\mu}\phi\right)_{n} - \left\{-\frac{r}{2}\left(\partial_{\mu}^{*}\partial_{\mu}^{}\phi\right)_{n} + P^{}\left(\phi_{n}\right)\right\}\right]\epsilon,
\end{align}
where $\epsilon$ is a constant Grassmann number.
In the presence of interactions, Eq.~(\ref{eq:3}) does not have the supersymmetry, but in the continuum limit, the supersymmetry has been perturbatively proven to restore in Ref.~\cite{Golterman:1988ta}.

In Eq.~(\ref{eq:3}) there are non-nearest neighbor hopping terms of $\phi$, which prevent us from building a tensor network on the square lattice.
To remove them, we insert two auxiliary scalar fields $G$ and $H$ into the boson part of the lattice action,
\begin{align}
  \label{eq:7}
  S_{\mathrm{B}}
  = \sum_{n}
  \biggl[&\frac{1}{2}\left(\partial_{\mu}^{}\phi\right)_{n}^{2}
           + \frac{1}{2}P^{}\left(\phi_{n}\right)^{2}
           - \frac{r}{2}\left(\partial_{\mu}^{*}\partial_{\mu}^{}\phi\right)_{n}P^{}\left(\phi_{n}\right)
           + \frac{1}{2}G_{n}^{2}
           + \frac{1}{2}H_{n}^{2} \nonumber\\
         &+ \sqrt{\frac{1-2r^{2}}{8}}G_{n}\left(\partial_{\mu}^{*}\partial_{\mu}^{}\phi\right)_{n}
           + \sqrt{\frac{1}{8}}H_{n}\left(-1\right)^{\delta_{\mu, 2}}\left(\partial_{\mu}^{*}\partial_{\mu}^{}\phi\right)_{n}\biggr].
\end{align}
In this manner, $G$ is decoupled with a certain value of the Wilson parameter, that is to say, $r = 1/\sqrt{2}$.
In the following subsections, we construct a tensor network representation of the partition function
\begin{align}
  \label{eq:8}
  Z
  = \int\mathcal{D}\phi\mathcal{D}H\mathcal{D}\psi e^{- S_{\mathrm{B}} - \frac{1}{2}\sum_{n}\bar{\psi}_{n}\left(D\psi\right)_{n}}
\end{align}
with a focus on the fermion and the boson part in order~\footnote{Note that the measure of the scalar field is divided by $\sqrt{2\pi}$ per site, and the measures of the antifermions do not exist in Eq.~(\ref{eq:8}) owing to the Majorana property of $\psi_{n}$.}.
For the boson part, we fix $r$ to $1/\sqrt{2}$ to deal with a single auxiliary field $H$.

\subsection{Tensor network for Pfaffian}

In this subsection, we construct a tensor network representation of the Pfaffian of the Majorana--Dirac operator.
The fundamental idea to obtain a tensor network representation for Dirac fermions is given in Refs.~\cite{Shimizu:2014uva,Takeda:2014vwa}.
We present a formulation for the Majorana case with the general Wilson parameter $r$ basically following them.

The Pfaffian is defined by
\begin{align}
  \label{eq:9}
  \mathrm{Pf}C^{*}D\left[\phi\right]
  = \int\mathcal{D}\psi e^{-\frac{1}{2}\sum_{n}\psi^{\mathrm{T}}_{n}\left(C^{*}D\psi\right)_{n}}.
\end{align}
In this report we use an explicit form of the gamma matrices and the charge conjugation matrix given by
\begin{align}
  \label{eq:1}
  \gamma_{1}=
  \begin{pmatrix}
    0 & 1 \\
    1 & 0
  \end{pmatrix}
        ,\quad \gamma_{2}=
        \begin{pmatrix}
          1 & 0 \\
          0 & -1
        \end{pmatrix}
              ,\quad C=
              \begin{pmatrix}
                0 & -1 \\
                1 & 0
              \end{pmatrix}
                    .
\end{align}
With this representation, the bilinear part in Eq.~(\ref{eq:9}) is written down as
\begin{align}
  \label{eq:10}
  -\frac{1}{2}\sum_{n}\psi_{n}^{\mathrm{T}}\left(C^{*}D\psi\right)_{n}
  = \sum_{n}
  \biggl[&\left(P^{\prime}\left(\phi\right) + 2r\right)\psi_{n,1}\psi_{n,2}
           + \frac{1+r}{4}\left(-\psi_{n+\hat{1}, 1} + \psi_{n+\hat{1}, 2}\right)\left(\psi_{n,1} + \psi_{n,2}\right) \nonumber\\
         &+ \frac{1-r}{4}\left(\psi_{n+\hat{1}, 1} + \psi_{n+\hat{1}, 2}\right)\left(-\psi_{n, 1} + \psi_{n, 2}\right) \nonumber\\
         &+ \frac{1+r}{2}\psi_{n+\hat{2},2}\psi_{n,1}
           + \frac{1-r}{2}\psi_{n+\hat{2}, 1}\psi_{n, 2}\biggr].
\end{align}
Thus the integrand in the RHS of Eq.~(\ref{eq:9}) turns out to be
\begin{align}
  \label{eq:11}
  e^{-\frac{1}{2}\sum_{n}\psi^{\mathrm{T}}_{n}\left(C^{*}D\psi\right)_{n}}
  = \sum_{\left\{x, t\right\}} \prod_{n}
  &\left[1 + \left(P^{\prime}\left(\phi\right) + 2r\right)\psi_{n,1}\psi_{n,2}\right] \nonumber\\
  &\cdot \left[\frac{1+r}{4}\left(-\psi_{n+\hat{1}, 1} + \psi_{n+\hat{1}, 2}\right)\left(\psi_{n,1} + \psi_{n,2}\right)\right]^{x_{n,1}} \nonumber\\
  &\cdot \left[\frac{1-r}{4}\left(\psi_{n+\hat{1}, 1} + \psi_{n+\hat{1}, 2}\right)\left(-\psi_{n, 1} + \psi_{n, 2}\right)\right]^{x_{n,2}} \nonumber\\
  &\cdot \left[\frac{1+r}{2}\psi_{n+\hat{2},2}\psi_{n,1}\right]^{t_{n,1}}
    \left[\frac{1-r}{2}\psi_{n+\hat{2}, 1}\psi_{n, 2}\right]^{t_{n,2}},
\end{align}
where the exponential functions have been expanded binomially using the nilpotency of Grassmann variables;
thus $x_{n, 1(2)}$ and $t_{n, 1(2)}$ run from 0 to 1, and $\left\{x, t\right\}$ represents $x_{n, 1(2)}$ and $t_{n, 1(2)}$ for all lattice sites $n$.
The second indices of $\psi$ denote the components in the spinor space.
Compared to the Dirac case, spinor components are completely mixed owing to the Majorana property of $\psi$.
However, the following procedure to make the tensor network is quite similar to the Dirac case: splitting the hopping factors and integrating out the original Grassmann variables $\psi$.
To complete the procedure, one needs to introduce new Grassmann variables for each direction.
As described in Ref.~\cite{Takeda:2014vwa}, we introduce $\left\{\eta, \xi\right\}$, and the definition of tensor is given by
\begin{align}
  \label{eq:13}
  &T^{\mathrm{\mathrm{F}}}_{x_{n,1}x_{n,2}t_{n,1}t_{n,2}x_{n-\hat{1},1}x_{n-\hat{1},2}t_{n-\hat{2},1}t_{n-\hat{2},2}}\left(\phi_{n}\right)
    \begin{aligned}[t]
      &\mathrm{d}\bar{\eta}_{n,2}^{x_{n,2}}\mathrm{d}\eta_{n,1}^{x_{n,1}}\mathrm{d}\bar{\xi}_{n,2}^{t_{n,2}}\mathrm{d}\xi_{n,1}^{t_{n,1}}\mathrm{d}\eta_{n,2}^{x_{n-\hat{1},2}}\mathrm{d}\bar{\eta}_{n,1}^{x_{n-\hat{1},1}}\mathrm{d}\xi_{n,2}^{t_{n-\hat{2},2}}\mathrm{d}\bar{\xi}_{n,1}^{t_{n-\hat{2},1}} \\
      &\cdot \left(\bar{\eta}_{n+\hat{1},1}\eta_{n,1}\right)^{x_{n,1}}
      \left(\bar{\eta}_{n,2}\eta_{n+\hat{1},2}\right)^{x_{n,2}}
      \left(\bar{\xi}_{n+\hat{2},1}\xi_{n,1}\right)^{t_{n,1}}
      \left(\bar{\xi}_{n,2}\xi_{n+\hat{2},2}\right)^{t_{n,2}}
    \end{aligned}\nonumber\\
  &=
    \begin{aligned}[t]
      \int&\mathrm{d}\psi_{n,1}\mathrm{d}\psi_{n,2} \left[1 + \left(P^{\prime}\left(\phi_{n}\right) + 2r\right)\psi_{n,1}\psi_{n,2}\right] \\
      &\cdot\left[\frac{\sqrt{1+r}}{2}\left(\psi_{n, 1} + \psi_{n, 2}\right)\mathrm{d}\eta_{n, 1}\right]^{x_{n,1}}
      \left[-\frac{\sqrt{1-r}}{2}\left(-\psi_{n, 1} + \psi_{n, 2}\right)\mathrm{d}\bar{\eta}_{n, 2}\right]^{x_{n,2}} \\
      &\cdot\left[\sqrt{\frac{1+r}{2}}\psi_{n, 1}\mathrm{d}\xi_{n, 1}\right]^{t_{n,1}}
      \left[-\sqrt{\frac{1-r}{2}}\psi_{n, 2}\mathrm{d}\bar{\xi}_{n, 2}\right]^{t_{n,2}}
      \left[\frac{\sqrt{1+r}}{2}\left(-\psi_{n, 1} + \psi_{n, 2}\right)\mathrm{d}\bar{\eta}_{n, 1}\right]^{x_{n-\hat{1},1}} \\
      &\cdot\left[\frac{\sqrt{1-r}}{2}\left(\psi_{n, 1} + \psi_{n, 2}\right)\mathrm{d}\eta_{n, 2}\right]^{x_{n-\hat{1},2}}
      \left[\sqrt{\frac{1+r}{2}}\psi_{n, 2}\mathrm{d}\bar{\xi}_{n, 1}\right]^{t_{n-\hat{2},1}}
      \left[\sqrt{\frac{1-r}{2}}\psi_{n, 1}\mathrm{d}\xi_{n, 2}\right]^{t_{n-\hat{2},2}} \\
      &\cdot \left(\bar{\eta}_{n+\hat{1}, 1}\eta_{n, 1}\right)^{x_{n,1}}
      \left(\bar{\eta}_{n, 2}\eta_{n+\hat{1}, 2}\right)^{x_{n,2}}
      \left(\bar{\xi}_{n+\hat{2}, 1}\xi_{n, 1}\right)^{t_{n,1}}
      \left(\bar{\xi}_{n, 2}\xi_{n+\hat{2}, 2}\right)^{t_{n,2}}.
    \end{aligned} \nonumber\\[-4ex]
\end{align}
We regard the LHS of Eq.~(\ref{eq:13}) as a tensor, which share the discrete d.o.f. ($x$ and $t$) with those of nearest neighbor sites,
and they construct a network.

Then the Pfaffian is represented as a product of tensors
\begin{align}
  \label{eq:15}
  \mathrm{Pf}C^{*}D\left[\phi\right]
  = \sum_{\left\{x, t\right\}} \int \prod_{n}
  &T^{\mathrm{\mathrm{F}}}_{x_{n,1}x_{n,2}t_{n,1}t_{n,2}x_{n-\hat{1},1}x_{n-\hat{1},2}t_{n-\hat{2},1}t_{n-\hat{2},2}}\left(\phi_{n}\right) \nonumber\\
  &\cdot \mathrm{d}\bar{\eta}_{n,2}^{x_{n,2}}\mathrm{d}\eta_{n,1}^{x_{n,1}}\mathrm{d}\bar{\xi}_{n,2}^{t_{n,2}}\mathrm{d}\xi_{n,1}^{t_{n,1}}\mathrm{d}\eta_{n,2}^{x_{n-\hat{1},2}}\mathrm{d}\bar{\eta}_{n,1}^{x_{n-\hat{1},1}}\mathrm{d}\xi_{n,2}^{t_{n-\hat{2},2}}\mathrm{d}\bar{\xi}_{n,1}^{t_{n-\hat{2}}} \nonumber\\
  &\cdot \left(\bar{\eta}_{n+\hat{1},1}\eta_{n,1}\right)^{x_{n,1}}
    \left(\bar{\eta}_{n,2}\eta_{n+\hat{1},2}\right)^{x_{n,2}}
    \left(\bar{\xi}_{n+\hat{2},1}\xi_{n,1}\right)^{t_{n,1}}
    \left(\bar{\xi}_{n,2}\xi_{n+\hat{2},2}\right)^{t_{n,2}}.
\end{align}

\subsection{Discretization of boson part and tensor network for total partition function}

In this subsection, we treat the boson part in Eq.~(\ref{eq:8}) with the fixed Wilson parameter $r = 1/\sqrt{2}$
\begin{align}
  \label{eq:16}
  e^{-S_{\mathrm{B}}}
  = \prod_{n} \prod_{\mu = 1}^{2} f_{\mu}\left(\phi_{n}, H_{n}; \phi_{n+\hat{\mu}}, H_{n+\hat{\mu}}\right),
\end{align}
where
\begin{align}
  \label{eq:22}
  &f_{\mu}\left(\phi_{n}, H_{n}; \phi_{n+\hat{\mu}}, H_{n+\hat{\mu}}\right) =
    \exp{
    \begin{aligned}[t]
      \Biggl[&- \frac{1}{4}\left(\phi_{n} - \phi_{n+\hat{\mu}}\right)^{2} - \frac{1}{2\sqrt{2}}\left(\phi_{n} - \phi_{n+\hat{\mu}}\right)P^{}\left(\phi_{n}\right) - \frac{1}{8}P^{}\left(\phi_{n}\right)^{2} \\
      &- \frac{1}{8}H_{n}^{2} + \sqrt{\frac{1}{8}}\left(-1\right)^{\delta_{\mu, 2}}\left(\phi_{n} - \phi_{n+\hat{\mu}}\right)H_{n} + \left(n \leftrightarrow n+\hat{\mu}\right)\Biggr].
    \end{aligned}
        }\nonumber\\[-5ex]
\end{align}
In a similar way to the fermion part, one has to expand $f$ with discrete indices which will be shared on each link in the network.
Actually, $f$ is a compact operator, and there are discrete spectra.
However, the scalar fields $\phi$ and $H$ are continuous, and this fact makes the numerical treatment hard.
To deal with this type of problem in the case of lattice $\phi^{4}$ theory, Shimizu presented a method to perform a numerical spectral decomposition of $f$ by using orthonormal functions~\cite{Shimizu:2012wfa}.
In this report, however, we discretize the scalar fields by approximating the integrals of them with the Gauss--Hermite quadrature
\begin{align}
  \label{eq:14}
  \int\mathrm{d}y e^{-y^{2}}g\left(y\right) \approx \sum_{\alpha=1}^{K} w_{\alpha} g\left(x_{\alpha}\right),
\end{align}
where $x_{\alpha}$ and $w_{\alpha}$ are the $\alpha$-th Gauss node and its weight of the Gauss--Hermite quadrature,
and $K$, the degree of the Hermite polynomial, determines the accuracy of this approximation for the arbitrary function $g\left(y\right)$.
The key point of this strategy is the presence of the damping factor in the LHS of Eq.~(\ref{eq:14}).
In $f$ there has to be the damping factor because the lattice action has the mass term.

Applying the Gauss--Hermite quadrature to the (path-)integral of $\phi$ and $H$, one obtains the discrete formula~\footnote{Of course, other types of numerical quadrature algorithms can be adopted here.}
:
\begin{align}
  \label{eq:17}
  &\int \mathrm{d}\phi_{n}\mathrm{d}H_{n} \prod_{\mu=1}^{2} f_{\mu}\left(\phi_{n-\hat{\mu}}, H_{n-\hat{\mu}}; \phi_{n}, H_{n}\right) f_{\mu}\left(\phi_{n}, H_{n}; \phi_{n+\hat{\mu}}, H_{n+\hat{\mu}}\right) \nonumber\\
  &\approx \sum_{\chi^{1}_{n}, \chi^{2}_{n} = 1}^{K} w_{\chi^{1}_{n}} w_{\chi^{2}_{n}} e^{x_{\chi^{1}_{n}}^{2} + x_{\chi^{2}_{n}}^{2}} \prod_{\mu=1}^{2} f_{\mu}\left(\phi_{n-\hat{\mu}}, H_{n-\hat{\mu}}; x_{\chi^{1}_{n}}, x_{\chi^{2}_{n}}\right) f_{\mu}\left(x_{\chi^{1}_{n}}, x_{\chi^{2}_{n}}; \phi_{n+\hat{\mu}}, H_{n+\hat{\mu}}\right),
\end{align}
where the same degree of the Hermite polynomial are used for both $\phi$ and $H$.
After that, $f$ is labeled by discrete indices, and one can numerically perform the singular value decomposition
\begin{align}
  \label{eq:18}
  &f_{1}\left(x_{\chi^{1}_{n}}, x_{\chi^{2}_{n}}; x_{\chi^{1}_{n+\hat{1}}}, x_{\chi^{2}_{n+\hat{1}}}\right)
    \approx \sum_{x_{n,\mathrm{b}} = 1}^{D_{\mathrm{B}}} U^{1}_{\chi_{n}, x_{n,\mathrm{b}}}\sigma^{1}_{x_{n,\mathrm{b}}}V^{1 \dagger}_{x_{n,\mathrm{b}}, \chi_{n+\hat{1}}}, \\
  &f_{2}\left(x_{\chi^{1}_{n}}, x_{\chi^{2}_{n}}; x_{\chi^{1}_{n+\hat{2}}}, x_{\chi^{2}_{n+\hat{2}}}\right)
    \approx \sum_{t_{n,\mathrm{b}} = 1}^{D_{\mathrm{B}}} U^{2}_{\chi_{n}, t_{n,\mathrm{b}}}\sigma^{2}_{t_{n,\mathrm{b}}}V^{2 \dagger}_{t_{n,\mathrm{b}}, \chi_{n+\hat{2}}},
\end{align}
where $\chi_{n}$ is defined by $\chi_{n} = \chi^{1}_{n} \otimes \chi^{2}_{n}$, and $D_{\mathrm{B}}$ is the dimension of new indices $x_{n, \mathrm{b}}$ and $t_{n, \mathrm{b}}$ which will become tensor indices.

Now, by combining the fermion and the boson part, the partition function can be expressed as a tensor network
\begin{align}
  \label{eq:19}
  Z
  &\approx \sum_{\left\{x, t\right\}} \int \sum_{\left\{\chi\right\}} \prod_{n}
    \begin{aligned}[t]
      &T^{\mathrm{\mathrm{F}}}_{x_{n,1}x_{n,2}t_{n,1}t_{n,2}x_{n-\hat{1},1}x_{n-\hat{1},2}t_{n-\hat{2},1}t_{n-\hat{2},2}}\left(x_{\chi^{1}_{n}}\right) w_{\chi^{1}_{n}} w_{\chi^{2}_{n}} e^{x_{\chi^{1}_{n}}^{2} + x_{\chi^{2}_{n}}^{2}}\\
      &\cdot U^{1}_{\chi_{n}, x_{n,\mathrm{b}}}\sqrt{\sigma^{1}_{x_{n,\mathrm{b}}}}
      U^{2}_{\chi_{n}, t_{n,\mathrm{b}}}\sqrt{\sigma^{2}_{t_{n,\mathrm{b}}}}
      \sqrt{\sigma^{1}_{x_{n-\hat{1},\mathrm{b}}}}V^{1 \dagger}_{x_{n-\hat{1},\mathrm{b}}, \chi_{n}}
      \sqrt{\sigma^{2}_{t_{n-\hat{2},\mathrm{b}}}}V^{2 \dagger}_{t_{n-\hat{2},\mathrm{b}}, \chi_{n}} \\
      &\cdot \mathrm{d}\bar{\eta}_{n,2}^{x_{n,2}}\mathrm{d}\eta_{n,1}^{x_{n,1}}\mathrm{d}\bar{\xi}_{n,2}^{t_{n,2}}\mathrm{d}\xi_{n,1}^{t_{n,1}}\mathrm{d}\eta_{n,2}^{x_{n-\hat{1},2}}\mathrm{d}\bar{\eta}_{n,1}^{x_{n-\hat{1},1}}\mathrm{d}\xi_{n,2}^{t_{n-\hat{2},2}}\mathrm{d}\bar{\xi}_{n,1}^{t_{n-\hat{2}}} \\
      &\cdot \left(\bar{\eta}_{n+\hat{1},1}\eta_{n,1}\right)^{x_{n,1}}
      \left(\bar{\eta}_{n,2}\eta_{n+\hat{1},2}\right)^{x_{n,2}}
      \left(\bar{\xi}_{n+\hat{2},1}\xi_{n,1}\right)^{t_{n,1}}
      \left(\bar{\xi}_{n,2}\xi_{n+\hat{2},2}\right)^{t_{n,2}}
    \end{aligned} \nonumber\\
  &= \sum_{\left\{x, t\right\}} \int \prod_{n}
    \begin{aligned}[t]
      &T_{(x_{n,1}, x_{n,2}, x_{n,\mathrm{b}})(t_{n,1}, t_{n,2}, t_{n,\mathrm{b}})(x_{n-\hat{1},1}, x_{n-\hat{1},2}, x_{n-\hat{1},\mathrm{b}})(t_{n-\hat{2},1}, t_{n-\hat{2},2}, t_{n-\hat{2},\mathrm{b}})} \\
      &\cdot \mathrm{d}\bar{\eta}_{n,2}^{x_{n,2}}\mathrm{d}\eta_{n,1}^{x_{n,1}}\mathrm{d}\bar{\xi}_{n,2}^{t_{n,2}}\mathrm{d}\xi_{n,1}^{t_{n,1}}\mathrm{d}\eta_{n,2}^{x_{n-\hat{1},2}}\mathrm{d}\bar{\eta}_{n,1}^{x_{n-\hat{1},1}}\mathrm{d}\xi_{n,2}^{t_{n-\hat{2},2}}\mathrm{d}\bar{\xi}_{n,1}^{t_{n-\hat{2}}} \\
      &\cdot \left(\bar{\eta}_{n+\hat{1},1}\eta_{n,1}\right)^{x_{n,1}}
      \left(\bar{\eta}_{n,2}\eta_{n+\hat{1},2}\right)^{x_{n,2}}
      \left(\bar{\xi}_{n+\hat{2},1}\xi_{n,1}\right)^{t_{n,1}}
      \left(\bar{\xi}_{n,2}\xi_{n+\hat{2},2}\right)^{t_{n,2}},
    \end{aligned}\nonumber\\[-4ex]
\end{align}
where
\begin{align}
  \label{eq:29}
  &T_{(x_{n,1}, x_{n,2}, x_{n,\mathrm{b}})(t_{n,1}, t_{n,2}, t_{n,\mathrm{b}})(x_{n-\hat{1},1}, x_{n-\hat{1},2}, x_{n-\hat{1},\mathrm{b}})(t_{n-\hat{2},1}, t_{n-\hat{2},2}, t_{n-\hat{2},\mathrm{b}})} \nonumber\\
  &= \sum_{\chi^{1}_{n}, \chi^{2}_{n}=1}^{K}
    \begin{aligned}[t]
      &T^{\mathrm{\mathrm{F}}}_{x_{n,1}x_{n,2}t_{n,1}t_{n,2}x_{n-\hat{1},1}x_{n-\hat{1},2}t_{n-\hat{2},1}t_{n-\hat{2},2}}\left(x_{\chi^{1}_{n}}\right) w_{\chi^{1}_{n}} w_{\chi^{2}_{n}} e^{x_{\chi^{1}_{n}}^{2} + x_{\chi^{2}_{n}}^{2}}\\
      &\cdot U^{1}_{\chi_{n}, x_{n,\mathrm{b}}}\sqrt{\sigma^{1}_{x_{n,\mathrm{b}}}}
      U^{2}_{\chi_{n}, t_{n,\mathrm{b}}}\sqrt{\sigma^{2}_{t_{n,\mathrm{b}}}}
      \sqrt{\sigma^{1}_{x_{n-\hat{1},\mathrm{b}}}}V^{1 \dagger}_{x_{n-\hat{1},\mathrm{b}}, \chi_{n}}
      \sqrt{\sigma^{2}_{t_{n-\hat{2},\mathrm{b}}}}V^{2 \dagger}_{t_{n-\hat{2},\mathrm{b}}, \chi_{n}}.
    \end{aligned}\nonumber\\[-4ex]
\end{align}
Note that the two types of approximation are introduced, the Gauss--Hermite quadrature for the integrals of scalar fields and the truncated singular value decomposition.
The tensor index $x_{n, 1(2)}$ and $x_{n, \mathrm{b}}$ run from 0 to 1 and from 1 to $D_{\mathrm{B}}$, respectively.
Then, the dimension of the integrated index $x_{n} = \left(x_{n, 1}, x_{n, 2}, x_{n, \mathrm{b}}\right)$ is $2\times 2\times D_{\mathrm{B}}$, and we call it as the bond dimension $D_{\mathrm{init}}$.
This notation is followed in next section.

\section{Numerical results}
\label{sec:Results}

\subsection{Partition functions of free Majorana--Wilson fermions}

\begin{figure}[htbp]
  \begin{minipage}{0.5\hsize}
    \centering
    \includegraphics[width=\hsize]{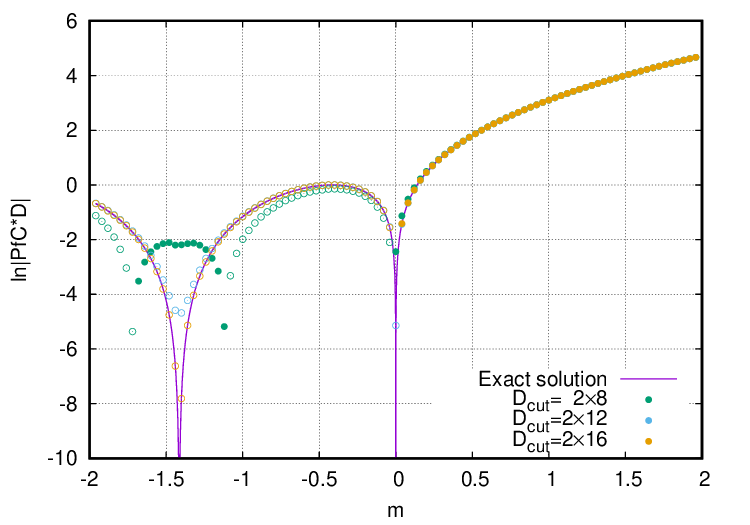}
    \caption{The partition function of two dimensional free Majorana--Wilson fermions as a function of $m$ on a $2\times 2$ lattice. The positive and negative sign of the partition function are represented as solid and open symbols, respectively. \label{fig:Z_2DfreeMajoranafermion_r0.7071067812_m-2.0-2.0_Dcut2x12_V2x2}}
  \end{minipage}
  \begin{minipage}{0.5\hsize}
    \centering
    \includegraphics[width=\hsize]{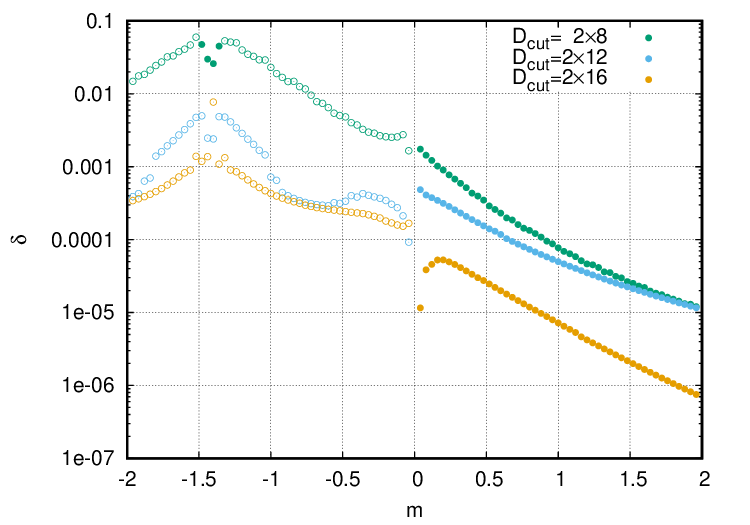}
    \caption{The relative error of the partition function of two dimensional free Majorana--Wilson fermions as a function of $m$ on a $32\times 32$ lattice. The positive and negative sign of the partition function are represented as solid and open symbols, respectively. \label{fig:Z_2DfreeMajoranafermion_r0.7071067812_m-2.0-2.0_Dcut2x12_V32x32}}
  \end{minipage}
\end{figure}

Figures~\ref{fig:Z_2DfreeMajoranafermion_r0.7071067812_m-2.0-2.0_Dcut2x12_V2x2} and ~\ref{fig:Z_2DfreeMajoranafermion_r0.7071067812_m-2.0-2.0_Dcut2x12_V32x32} show the partition function of free Majorana--Wilson fermions, which are computed by the TRG with a fixed bond dimension $D_{\mathrm{cut}}$ and periodic boundary conditions.
Specifically, we compute Eq.~(\ref{eq:9}) with
\begin{align}
  D = \partial\hspace{-1.5mm}/^{\mathrm{s}}
  - \frac{r}{2}\partial^{*}_{\mu}\partial^{}_{\mu}
  + m.
\end{align}
To compute it, we use the Grassmann TRG~\cite{Gu:2010yh,Gu:2013gba}, which has been extended to relativistic fermion systems in Refs.~\cite{Shimizu:2014uva,Takeda:2014vwa}.
We follow Ref.~\cite{Takeda:2014vwa} for the coarse-graining procedure except for the definition of $D_{\mathrm{cut}}$.
It is defined as the dimension of $x_{n}=\left(x_{n, \mathrm{f}}, x_{n, \mathrm{b}}\right)$ in this report, on the other hand, it is defined as the dimension of $x_{n, \mathrm{b}}$ in Ref.~\cite{Takeda:2014vwa}, where $x_{n, \mathrm{f} (\mathrm{b})}$ represents the fermionic (bosonic) index of tensor (see Ref.~\cite{Takeda:2014vwa}).

The results for $2\times 2$ and $32\times 32$ space-time lattice are shown in Figs.~\ref{fig:Z_2DfreeMajoranafermion_r0.7071067812_m-2.0-2.0_Dcut2x12_V2x2} and ~\ref{fig:Z_2DfreeMajoranafermion_r0.7071067812_m-2.0-2.0_Dcut2x12_V32x32}, respectively,
and for $32\times 32$ case, the relative error is shown.
The behavior of numerical errors is qualitatively the same in each case.
From the viewpoint of data compression, the larger $D_{\mathrm{cut}}$ contains the more information, and the results of larger $D_{\mathrm{cut}}$ provides the more accurate results as expected.
There are fermion zero modes at $m=0.0$ and $m=-\sqrt{2}$, and the error grows near such points.
The Pfaffian flips its sign depending on the value of $m$, and this reflects the fact that the Pfaffian is not positive definite.

\subsection{Witten index of free Wess--Zumino model}

\begin{figure}[htbp]
  \begin{minipage}{0.5\hsize}
    \centering
    \includegraphics[width=\hsize]{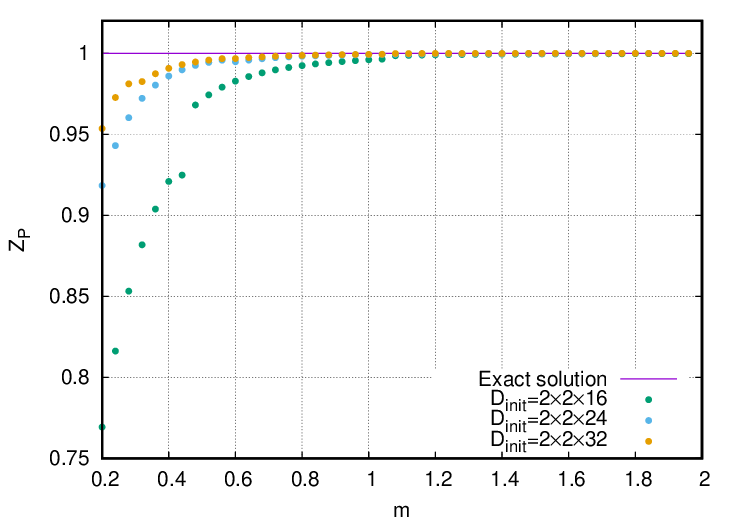}
    \caption{The Witten index for the free case as a function of $m$ on a $2 \times 2$ lattice. \label{fig:Z_2DfreeWZ_m0.0-2.0_Kphi64_KH64_Dinit2x2x16-2x2x32_V2x2}}
  \end{minipage}
  \begin{minipage}{0.5\hsize}
    \centering
    \includegraphics[width=\hsize]{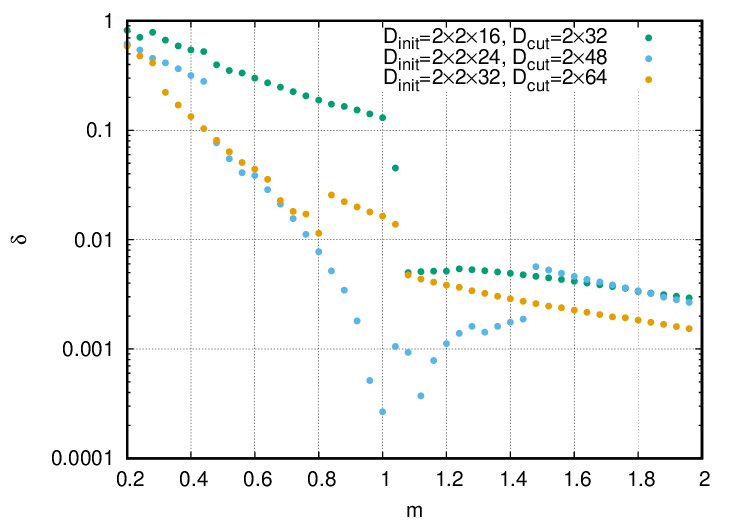}
    \caption{The relative error of the Witten index for the free case as a function of $m$ on a $8 \times 8$ lattice. \label{fig:Z_2DfreeWZ_m0.0-2.0_Kphi64_KH64_Dinit2x2x16-2x2x32_Dcut2x32-2x64_V8x8}}
  \end{minipage}
\end{figure}

Figures~\ref{fig:Z_2DfreeWZ_m0.0-2.0_Kphi64_KH64_Dinit2x2x16-2x2x32_V2x2} and~\ref{fig:Z_2DfreeWZ_m0.0-2.0_Kphi64_KH64_Dinit2x2x16-2x2x32_Dcut2x32-2x64_V8x8} show the partition functions of the free Wess--Zumino model whose superpotential is given by
\begin{align}
  \label{eq:20}
  W^{\mathrm{free}}\left(\phi\right)
  = \frac{1}{2}m\phi^{2}.
\end{align}
The partition functions are computed with periodic boundary conditions and called the Witten index which is regarded as an indicator of the supersymmetry breaking~\cite{Witten:1982df}.
The relative error is shown in $8\times 8$ space-time volume case.
In this subsection, we fix the degree of the Hermite polynomial for both $\phi$ and $H$ as $K=64$.
The tensor in Eq.~(\ref{eq:19}) has $D_{\mathrm{init}}^{4} = \left(2 \times 2 \times D_{\mathrm{B}}\right)^{4}$ components at the initial stage, and it can grow in coarse-graining steps.
We apply the GTRG also in this subsection and cut the growth of the bond dimension by $D_{\mathrm{cut}}$.

In the case of the free Wess--Zumino model, the exact solution of the Witten index is known to be exactly one even on the lattice, and one can see that the larger bond dimension provides the more accurate results.
Thus we can conclude that the tensor network representation in Eq.~(\ref{eq:19}) is correct.
The reason of low accuracy in the small $m$ region is vanishing of the mass term.
This means vanishing of the damping factor in the LHS of Eq.~(\ref{eq:14}).
If one thinks of the interacting case, the $\phi^{4}$ interaction term guarantees the fast damping of $f$, and such a bad property will go away.

\section{Summary and outlook}
\label{sec:Summary}

We have constructed a tensor network representation of the two dimensional lattice $\mathcal{N}=1$ Wess--Zumino model, and the correctness of the new formulation is confirmed numerically.
Now we are ready to turn to the interacting case whose superpotential is given by
\begin{align}
  \label{eq:21}
  W^{\mathrm{interaction}}\left(\phi\right)
  = \frac{1}{3}g\phi^{3}
  - \frac{m^{2}}{4g}\phi,
\end{align}
where $g$ is the coupling constant.
The model is known to exhibit the spontaneous supersymmetry breaking in the presence of interaction.
To check whether the phase structure is consistent with previous results in Ref.~\cite{Steinhauer:2014yaa} or not, we can compute the vacuum expectation value of the scalar field or fermionic/bosonic Green's functions using the methods described in Ref.~\cite{2008PhRvB..78t5116G}.
We are now working in this direction.

Another possible approach to supersymmetric lattice field theories is the domain-wall discretization for fermions and bosons.
The coarse-graining technique for higher dimensional fermion systems is invented in Ref.~\cite{Sakai:2017jwp}, and the computational cost might be reasonable in three dimensions.

\section*{Acknowledgments}
We thank Dr. Y. Shimizu for helpful discussions.
This work is supported in part by
JSPS KAKENHI Grant Numbers JP16K05328, JP17K05411,
Grants-in-Aid for Scientific Research from the Ministry of Education, Culture, Sports, Science and Technology (MEXT) (Nos. 15H03651),
MEXT as ``Exploratory Challenge on Post-K computer (Frontiers of Basic Science: Challenging the Limits)'',
and the MEXT-Supported Program for the Strategic Research Foundation at Private Universities Topological Science (Grant No. S1511006).

\bibliography{References}

\end{document}